\documentclass[aps,pre,floatfix,nofootinbib,showpacs,groupedaddress,superscriptaddress]{revtex4}
\usepackage[sort&compress]{natbib}
\usepackage{graphicx}
\usepackage{epsfig}

\begin{document}

\title{Fluctuations in Gene Regulatory Networks as Gaussian Colored Noise}
\author{Ming-Chang Huang}
\affiliation{Department of Physics and Center for Nonlinear and Complex Systems,
Chung-Yuan Christian University, Chungli, 32023 Taiwan}
\author{Jinn-Wen Wu}
\affiliation{Department of Applied Mathematics, Chung-Yuan Christian University, Chungli,
32023 Taiwan}
\author{Yu-Pin Luo}
\affiliation{Department of Electronic Engineering, National Formosa University, Yunlin
County 632, Taiwan}
\author{Karen G. Petrosyan}
\affiliation{Institute of Physics, Academia Sinica, Nankang, Taipei 11529, Taiwan}
\date{\today }

\begin{abstract}
The study of fluctuations in gene regulatory networks is extended to the
case of Gaussian colored noise. Firstly, the solution of the corresponding
Langevin equation with colored noise is expressed in terms of an Ito integral.
Then, two important lemmas concerning the variance of an Ito integral
and the covariance of two Ito integrals are shown. Based on the lemmas, we
give the general formulae for the variances and covariance of molecular
concentrations for a regulatory network near a stable equilibrium explicitly.
Two examples, the gene auto-regulatory network and the toggle switch,
are presented in details. In general, it is found that the finite correlation time
of noise reduces the fluctuations and enhances the correlation between the
fluctuations of the molecular components.
\end{abstract}

\maketitle

\section{Introduction}

A regulatory network of gene expression consists of a group of genes which
co-regulate one another's expression. The networks provide a fundamental
description of cellular function that is the subject of the recently emerged
field of systems biology \cite{alon}. The advance of experimental techniques
in constructing synthetic networks provided with the basic elements, such as
a switch \cite{ptashne,hasty,gardner} and an oscillator \cite%
{elowitz,atkinson}, for the design of biological circuits \cite{alon}. For
such elements, the characteristic features are determined mainly by the
regulation scheme, and fluctuations always appear to be significant due to
low copy numbers of molecules and stochastic nature of biochemical reactions
\cite{rao,pedraza}. In clarifying the nature of regulation mechanism, one of the
important questions is to understand the way of accommodating the
fluctuations meanwhile maintaining the stability for a genetic network.
Recently, significant progress has been made along this aspect. One of the
noticeable examples is the auto-regulatory network of a single gene for
which, the protein encoded in the gene serves as the regulator of itself
through either negative or positive feedback. Such autoregulation is an
ubiquitous motif in biochemical pathways \cite{alon}. It was demonstrated by
Becskei and Serrano that an autoregulatory network with negative feedback
may gain stability \cite{becskei}. Further analysis given by Thattai and van
Oudenaarden \cite{thattai} and by Ozbudak et al. \cite{ozbudak} indicate
that noise is essentially determined at the translational level and negative
feedback can suppress the intrinsic noise. Another example is the genetic
toggle switch. Such a switch consists of two transcription factors which
regulate each other's synthesis negatively \cite{gardner}. Detailed analysis
given by Cherry and Adler \cite{cherry} shows that the cooperative binding
of two or more proteins in negative regulation is required, in general, for
a switch to have two distinct stable states. Further study carried out by Warren
and ten Wolde \cite{warren} indicates that the switch stability can be enhanced
by overlapping the upperstream regulatory domains such that competing
regulatory molecules mutually exclude each other, and robustness against
biochemical noise may provide a selection pressure that drives operons
together in the course of evolution. Recently, the results reported by
Loinger et al \cite{loinger} reveal that a suitable combination of network
structure and stochastic effects gives rise to bistability even without
cooperative binding.

Stochastic fluctuations associated with a system are often assumed to be
Gaussian white noise in nature. However, the zero correlation time for white
noise assumes an infinite relaxation time. Thus, it is important to
incorporate the effect of finite correlation time of noise into the study of
stochastic fluctuations. As the treatment for the effect induced by the Gaussian
colored noise in a regulatory network being still lacking in the literature,
we intend to fill up this gap in this paper. In modelling the dynamics of a
regulatory network, rate-equation approach is often used; the approach
reflects the macroscopic observation of deterministic nature. Noise-induced
effects may be incorporated into the framework by employing the master
equation and then proceeding via stochastic Monte Carlo simulations \cite%
{gillespie} as was done for the first time for gene regulatory networks by
Arkin et al \cite{arkin}. In general, master equations are discrete in
nature. By using the technique of $\Omega $-expansion \cite{kampen}, one can
convert a master equation to continuous Fokker-Planck equation which then
can be treated analytically by various approximations. Based on the
Fokker-Planck equation, Tao and Tao et al. derived the corresponding linear
noise Fokker-Planck equation which is suitable for the study of fluctuations
caused by the white noise \cite{tao,tao-c}. One may include the Gaussian colored
noise into the frame of Fokker-Planck equation by using the scheme of
unified colored-noise approximation \cite{jung} or via kinetic Monte-Carlo
simulations as was implemented for gene regulatory networks in \cite%
{shahrezaei}. Alternatively, we first establish the equivalent Langevin
description for a system with colored noise in the linearized region of a
stable point \cite{gardiner}. The solution of the corresponding Langevin
equation is given in the form of an Ito integral, and the fluctuations of
molecular concentrations then can be evaluated by using the two lemmas,
which concern with the variances and covariances of Ito integrals, shown in
this work. As the correlation time of noise is set to vanish, we recover the
results of fluctuations for white noise. Based on this approach, we analyze
the stochastic fluctuations of autoregulatory networks and toggle switches
with general cooperative binding, and the results are found to be in good
agreement with those obtained from numerical simulations. In general, the
appearance of finite correlation time of noise decreases the fluctuations of
a system and enhances the correlation between fluctuations.

This paper is organized as follows. In section II, we first set up the model
system for the study of the steady-state statistics of a regulatory network
near a stable equilibrium. The rate equations are first
linearized about the stable equilibrium. Starting with the rate equations,
we then use the technique of $\Omega $-expansion to obtain the corresponding
Fokker-Planck equation of the system. In section III, the equivalent
Langevin description of the Fokker-Planck equation is given, and the
solution of the Langevin equation with colored noise is expressed in terms
of an Ito integral. Then, two important lemmas concerning the variance of
an Ito integral and the covariance of two Ito integrals are shown. Based on
the lemmas, we obtain the variances and covariance of molecular
concentrations for a two-dimensional regulatory network near a stable
equilibrium. In sections IV, the formulae resulting from the lemmas are
applied to auto-regulatory networks and toggle switches to study the
stochastic fluctuations of the systems. The comparison between the results
and numerical simulations is given. In particular, the effects of noise
correlation time on the amount of fluctuations and on the correlation
between the two molecular components are analyzed. Finally, we summarize the
results in section V.

\section{Fokker-Planck equation}

Consider a two-dimensional regulatory network of gene expression defined by
the macroscopic rate equation,

\begin{equation}
dx\left( t\right) = f\left( x\right) dt,  \label{eq01}
\end{equation}
where the state variables $x^{\tau }\left( t\right) =\left( x_{1}\left(
t\right) ,x_{2}\left( t\right) \right) $ are molecular concentrations, and
the forces $f^{\tau }\left( x\right) =\left( f_{1}\left( x\right)
,f_{2}\left( x\right) \right) $ determine the time evolutions of the state
variables. Here, the superscript $\tau $ denotes the transpose of a vector.
A stable point $x^{\ast }$, specified by zero force $f\left( x^{\ast
}\right) =0$, represents a stable stationary state. For the region near the
stable equilibrium, the leading order of the drift force in $x$ gives
\begin{equation}
f\left( y\right) = F\left( x^{\ast }\right) \cdot y,  \label{eq02}
\end{equation}
where the stable point is chosen as the origin, $y=x-x^{\ast }$, in our
two-dimensional space, and the elements of $F\left( x^{\ast }\right) $ are
defined as $F_{ij}\left( x^{\ast }\right) =\left. \partial f_{i}\left(
x\right) /\partial x_{j}\right\vert _{x=x^{\ast }}$. From now on, we drop
the arguments whenever the matrix elements are understood as functions of
the equilibrium stable point $x^{\ast }$. The matrix $F$ can be diagonalized
by means of the transformation matrix $P$,

\begin{equation}
F=P\cdot \left(
\begin{array}{cc}
-\lambda _{1} & 0 \\
0 & -\lambda _{2}%
\end{array}%
\right) \cdot P^{-1},  \label{eq03}
\end{equation}
where $\lambda _{1},\lambda _{2}>0$ for the stable point, and $\det \left(
P\right) = 1$. The matrix $P$ is set as
\begin{equation}
P=\left(
\begin{array}{cc}
p_{11} & p_{12} \\
p_{21} & p_{22}%
\end{array}%
\right) ,  \label{eq04}
\end{equation}%
and \ this gives the inverse as
\begin{equation}
P^{-1}=\left(
\begin{array}{cc}
p_{22} & -p_{12} \\
-p_{21} & p_{11}%
\end{array}%
\right) .  \label{eq05}
\end{equation}

In general, the drift force of Eq. (\ref{eq01}) can be expressed as the sum
of two terms,
\begin{equation}
f\left( x\right) =R\left( x\right) -\Theta \cdot x,  \label{eq006}
\end{equation}%
where the functions, $R^{\tau }\left( x\right) =\left( R_{1}\left(
x\right),R_{2}\left( x\right) \right) $, describe the synthesis or feedback
regulation of molecule, and $\Theta $ is a $2\times 2$ constant matrix with
the elements given by $\Theta _{ij}=\delta _{ij}\theta ^{\left( i\right) }$
with the degradation rate $\theta ^{\left( i\right) }$ for molecular
concentration $x_{i}$. Since the synthesis or feedback regulation of
molecule concentration $x_{i}$ depends only on the concentration of other
component, we have $R_{i}\left( x\right) =R_{i}\left( x_{j}\right) $ for $%
j\neq i$. The stochastic fluctuations can be incorporated into Eq. (\ref%
{eq01}) by means of the master equation approach. For this, we introduce the
volume factor $\Omega $ to relate the molecular numbers $n^{\tau }=\left(
n_{1}\text{, }n_{2}\right) $ to the concentrations $x^{\tau }=\left(
x_{1},x_{2}\right) $ as $n^{\tau }=\Omega x^{\tau }$. In terms of molecular
numbers $n$, the corresponding master equation for Eq. (\ref{eq01}) with the
drift force given by Eq. (\ref{eq006}) is
\begin{equation}
\frac{\partial P\left( n,t\right) }{\partial t}=\sum_{i=1}^{2}\left(
E_{i+}-1\right) \left[ \left( \theta ^{\left( i\right) }n_{i}\right) P\left(
n,t\right) \right] +\Omega \sum_{i=1}^{2}R_{i}\left( x\right) \left[ E_{i-}-1%
\right] P\left( n,t\right) ,  \label{eq007}
\end{equation}
where $P\left( n,t\right) $ is the probability distribution, the step
operators $E_{i\pm }$ are defined as $E_{i\pm }G\left( n_{i}\right) =G\left(
n_{i}\pm 1\right) $ for a function of molecular numbers $G\left( n\right) $,
and the fact, $R_{i}\left( x\right) =R_{i}\left( x_{j}\right) $ for $j\neq i$%
, is used for the term in the second sum. Then, the technique of $\Omega $%
-expansion \cite{kampen} is employed to transfer the discrete process of Eq.
(\ref{eq007}) to a continuous process described by the Fokker-Planck
equation,
\begin{equation}
\frac{\partial \rho \left( x,t\right) }{\partial t}+\nabla \cdot j\left(
x,t\right) =0,  \label{eq008}
\end{equation}%
where $\left( \nabla \right) ^{\tau }=\left( \partial /\partial x_{1}\text{,
}\partial /\partial x_{2}\right) $, $\rho \left( x,t\right) $ is the
distribution density, and $j\left( x,t\right) $ is the density current given
as \
\begin{equation}
j\left( x,t\right) =f\left( x\right) \rho \left( x,t\right) -\left[ D\left(
x\right) \cdot \nabla \right] \rho \left( x,t\right) .  \label{eq009}
\end{equation}
Based on Eq. (\ref{eq007}), we obtain the diffusion matrix $D\left( x\right)
$ of Eq. (\ref{eq009}) as $D_{ij}\left( x\right) =\delta _{ij}d^{\left(
i\right) }\left( x\right) $, which takes the diagonal form with the diagonal
elements given as
\begin{equation}
d^{\left( i\right) }\left( x\right) =\frac{1}{2\Omega }\left[ R_{i}\left(
x\right) +\theta ^{\left( i\right) }x_{i}\right] .  \label{eq010}
\end{equation}

For the linear region specified by Eq. (\ref{eq02}) we obtain the
corresponding Fokker-Planck equation by expanding the density current $%
J\left( x,t\right) $ of Eq. (\ref{eq009}) around the stable point $x^{\ast }$%
. The result reads
\begin{equation}
\frac{\partial \rho _{L}\left( y,t\right) }{\partial t}+\nabla \cdot
J_{L}\left( y,t\right) =0,  \label{eq011}
\end{equation}%
where $J_{L}\left( y,t\right) $ contains only the leading order terms of $%
J\left( x,t\right) $ in $1/\Omega $,
\begin{equation}
J_{L}\left( y,t\right) =\left[ F\left( x^{\ast }\right) \cdot y-D\left(
x^{\ast }\right) \cdot \nabla \right] \rho _{L}\left( y,t\right) .
\label{eq012}
\end{equation}
This leads to an Ornstein-Uhlenbeck process \cite{kampen,gardiner,uhlenbeck}
in which, the drift force is linear and the diffusion is given by a constant
matrix.

\section{Equivalent Langevin description}

For the Fokker-Planck equation of Eq. (\ref{eq011}), we have the equivalent
Langevin description specified by the stochastic differential equation,
\begin{equation}
dy\left( t\right) =F\cdot y\left( t\right) dt+\eta \left( t\right) dt.
\label{eq05-1}
\end{equation}
From hereafter the stochastic fluctuations, described by the variables $\eta
^{\tau }\left( t\right) =\left( \eta _{1}\left( t\right) ,\eta _{2}\left(
t\right) \right) $, will be assumed to be Gaussian colored noises. The two
independent colored noises are specified by the differential equation,
\begin{equation}
d\eta \left( t\right) =-\Gamma \cdot \eta \left( t\right) dt+\Gamma \cdot
\Lambda \cdot dW\left( t\right) ,  \label{eq05-2}
\end{equation}
where the constant matrices, $\Gamma $ and $\Lambda $, and the vector $%
dW\left( t\right) $ are defined as follows. The elements of the $\Gamma $
matrix are given as $\Gamma _{ij}=\delta _{ij}\left( 1/\tau ^{\left(
i\right) }\right) $, where $\tau ^{\left( i\right) }$ is the
correlation time of the noise $\eta _{i}$ and the Kronecker delta $\delta
_{i,j}$ is equal to $1$ for $i=j$ and $0$ otherwise. The $\Lambda $ matrix
is related to the diffusion matrix evaluated at the stable point, $D=D\left(
x^{\ast }\right) $, by $\Lambda \cdot \Lambda =2D$. Based on the form of the
diffusion matrix of Eq. (\ref{eq010}), we have $\Lambda _{ij}=\delta _{ij}%
\sqrt{2d^{\left( i\right) }\left( x^{\ast }\right) }$ for the elements of $%
\Lambda $. Furthermore, the variables $dW^{\tau }\left( t\right) =\left(
dw_{1}\left( t\right) ,dw_{2}\left( t\right) \right) $ describe two
independent Wiener processes. By rewriting $dW\left( t\right) \rightarrow
\xi \left( t\right) dt$, the conditions, $\left\langle \xi \left( t\right)
\right\rangle =0$ and $\left\langle \xi \left( t\right) \xi ^{\tau }\left(
s\right) \right\rangle =\delta \left( t-s\right) $ with the Dirac delta
function $\delta \left( t\right) $, specify the Gaussian white noise.

The solution of Eq. (\ref{eq05-2}) can be expressed in terms of Ito integral
as
\begin{equation}
\eta \left( t\right) =\eta \left( 0\right) \exp \left( -t\Gamma \right)
+\int_{0}^{t}\left[ \exp \left( -\left( t-s\right) \Gamma \right) \right]
\cdot \Gamma \cdot \Lambda \cdot dW\left( s\right) .  \label{eq05-3}
\end{equation}
This yields the correlation function of the Gaussian colored noise as
\begin{equation}
\left\langle \eta \left( t\right) \eta ^{\tau }\left( t^{\prime }\right)
\right\rangle =\Gamma \cdot D\exp \left( -\left\vert t-t^{\prime
}\right\vert \Gamma \right) .  \label{eq05-4}
\end{equation}
Then the solution of Eq. (\ref{eq05-1}) can be written as

\begin{eqnarray}
y\left( t\right) = y\left( 0\right) \exp \left( tF\right) + \int_{0}^{t}
\left[ \exp \left( t-u\right) F\right] \cdot \left\{ \eta \left( 0\right)
\exp \left( -u\Gamma \right) +\int_{0}^{u}\left[ \exp \left( -\left(
u-s\right) \Gamma \right) \right] \cdot \Gamma \cdot \Lambda \cdot dW\left(
s\right) \right\} du.  \label{eq05-5}
\end{eqnarray}
Since we are interested in the fluctuations around the stable point $x^{\ast
} $, only the asymptotic behavior of the solution matters, and we may set
the initial condition as $y\left( 0\right) =0$ and $\eta \left( 0\right) =0$
to obtain
\begin{equation}
y\left( t\right) =\int_{0}^{t}\left[ \exp \left( t-u\right) F\right] \cdot
\left\{ \int_{0}^{u}\left[ \exp \left( -\left( u-s\right) \Gamma \right) %
\right] \cdot \Gamma \cdot \Lambda \cdot dW\left( s\right) \right\} du.
\label{eq05-5-1}
\end{equation}
Moreover, the order of the double integrations can be changed properly to
yield
\begin{equation}
y\left( t\right) =\int_{0}^{t}\left\{ \int_{s}^{t}\left[ \exp \left(
t-u\right) F\right] \cdot \left[ \exp \left( -\left( u-s\right) \Gamma
\right) \right] \cdot \Gamma \cdot \Lambda du\right\} \cdot dW\left(
s\right) .  \label{eq05-6}
\end{equation}%
Thus, the form of the solution, in general, is an Ito integral.

Two important lemmas for the evaluations of variances and covariances of Ito
integrals are shown as follows. Consider an Ito integral in the form of
\begin{equation}
I\left( t\right) =\int_{0}^{t}V\left( t,s\right) dw\left( s\right) ,
\label{eq08}
\end{equation}
for a random Wiener process $dw$. The integral can be expressed as the
discrete form,
\begin{equation}
I\left( t\right) =\lim_{N\rightarrow \infty }\sum_{k=0}^{N-1}V\left(
t,s_{k}\right) \left[ w\left( s_{k+1}\right) -w\left( s_{k}\right) \right] ,
\label{eq08-1}
\end{equation}
with $s_{0}=0$ and $s_{k}=k\left( t/N\right) $. This leads to the variance
of $I\left( t\right) $ as
\begin{equation}
\sigma ^{2}\left[ I\left( t\right) \right] =\lim_{N\rightarrow \infty
}\sum_{k=0}^{N-1}V^{2}\left( t,s_{k}\right) \sigma ^{2}\left[ w\left(
s_{k+1}\right) -w\left( s_{k}\right) \right] .  \label{eq09}
\end{equation}
Here the variance of $I\left( t\right) $ is defined as $\sigma ^{2}\left[
I\left( t\right) \right] =\left\langle I^{2}\left( t\right) \right\rangle
-\left\langle I\left( t\right) \right\rangle ^{2}$ with the expectation
value $\left\langle \cdot \right\rangle $ taken with respect to the
distribution of noise at time $t$. Based on the equality for a Wiener
process, $\sigma ^{2}\left[ w\left( s_{k+1}\right) -w\left( s_{k}\right) %
\right] =s_{k+1}-s_{k}$, we can express Eq. (\ref{eq09}) as the integral
form and obtain the first lemma, namely, the variance of $I\left( t\right) $
is
\begin{equation}
\sigma ^{2}\left[ I\left( t\right) \right] =\int_{0}^{t}V^{2}\left(
t,s\right) ds.  \label{eq11}
\end{equation}

The first lemma can be further extended to the case of two dimensions in a
straightforward way. Consider the Ito integrals, $J^{\tau }\left( t\right)
=\left( J_{1}\left( t\right) ,J_{2}\left( x\right) \right) $, defined as
\begin{equation}
J\left( t\right) =\int_{0}^{t}M\left( t,s\right) \cdot dW\left( s\right) ,
\label{eq12}
\end{equation}
where $M\left( t,s\right) $ is a $2\times 2$ matrix,
\begin{equation}
M\left( t,s\right) =\left(
\begin{array}{cc}
m_{11}\left( t,s\right) & m_{12}\left( t,s\right) \\
m_{21}\left( t,s\right) & m_{22}\left( t,s\right)%
\end{array}%
\right) ,  \label{eq21-1}
\end{equation}
and the two variables $dw_{1}$ and $dw_{2}$ in $dW^{\tau }\left( t\right)
=\left( dw_{1}\left( t\right) ,dw_{2}\left( t\right) \right) $ describe two
independent random Wiener processes. Following the first lemma along with
the property $\left\langle dW\left( t\right) dW^{\tau }\left( s\right)
\right\rangle =\delta \left( t-s\right) dt$, one can show that the variances
of $J_{1}\left( t\right) $ and $J_{2}\left( t\right) $ are
\begin{equation}
\sigma ^{2}\left[ J_{i}\left( t\right) \right] =\int_{0}^{t}\left[
m_{i1}^{2}\left( t,s\right) +m_{i2}^{2}\left( t,s\right) \right] ds
\label{eq13}
\end{equation}
for $i=1,2$, and the covariance is
\begin{equation}
E\left[ J_{1}\left( t\right) J_{2}\left( t\right) \right] =\int_{0}^{t}\left[
m_{11}\left( t,s\right) m_{21}\left( t,s\right) +m_{12}\left( t,s\right)
m_{22}\left( t,s\right) \right] ds.  \label{eq14}
\end{equation}
Here the covariance of $J_{1}\left( t\right) $ and $J_{2}\left( t\right) $
is defined as $E\left[ J_{1}\left( t\right) J_{2}\left( t\right) \right]
=\left\langle J_{1}\left( t\right) J_{2}\left( t\right) \right\rangle
-\left\langle J_{1}\left( t\right) \right\rangle \left\langle J_{2}\left(
t\right) \right\rangle $. These constitute the second lemma.

The second lemma, Eqs. (\ref{eq13}) and (\ref{eq14}), with the limit $%
t\rightarrow \infty $ can be employed directly to determine the variances
and covariance of $y\left( t\right) $ of Eq. (\ref{eq05-6}) near the stable
equilibrium $y=0$. The results thus obtained are summarized in the
following. The variances of $y_{1}$ and $y_{2}$, referred as $\sigma
_{1}^{2} $ and $\sigma _{2}^{2}$ respectively, are
\begin{equation}
\sigma _{1}^{2}\left( \tau ^{\left( 1\right) },\tau ^{\left( 2\right)
}\right) =d^{\left( 1\right) }\left[ \frac{q_{3}^{2}}{\lambda _{1}c_{11}}+%
\frac{q_{4}^{2}}{\lambda _{2}c_{21}}-\frac{2q_{3}q_{4}\widetilde{c}_{1}}{%
\lambda _{1}+\lambda _{2}}\right] +d^{\left( 2\right) }\left[ \frac{q_{1}^{2}%
}{\lambda _{1}c_{12}}+\frac{q_{1}^{2}}{\lambda _{2}c_{22}}-\frac{2q_{1}^{2}%
\widetilde{c}_{2}}{\lambda _{1}+\lambda _{2}}\right] ,  \label{eq24}
\end{equation}
and
\begin{equation}
\sigma _{2}^{2}\left( \tau ^{\left( 1\right) },\tau ^{\left( 2\right)
}\right) =d^{\left( 1\right) }\left[ \frac{q_{2}^{2}}{\lambda _{1}c_{11}}+%
\frac{q_{2}^{2}}{\lambda _{2}c_{21}}-\frac{2q_{2}^{2}\widetilde{c}_{1}}{%
\lambda _{1}+\lambda _{2}}\right] +d^{\left( 2\right) }\left[ \frac{q_{3}^{2}%
}{\lambda _{2}c_{22}}+\frac{q_{4}^{2}}{\lambda _{1}c_{12}}-\frac{2q_{3}q_{4}%
\widetilde{c}_{2}}{\lambda _{1}+\lambda _{2}}\right] ;  \label{eq25}
\end{equation}
and the covariance between $y_{1}$ and $y_{2}$, denoted as $E$, is
\begin{equation}
E\left( \tau ^{\left( 1\right) },\tau ^{\left( 2\right) }\right) =d^{\left(
1\right) }\left[ \frac{q_{2}q_{3}}{\lambda _{1}c_{11}}+\frac{q_{2}q_{4}}{%
\lambda _{2}c_{21}}-\frac{\left( q_{2}q_{3}+q_{2}q_{4}\right) \widetilde{c}%
_{1}}{\lambda _{1}+\lambda _{2}}\right] +d^{\left( 2\right) }\left[ \frac{%
q_{1}q_{4}}{\lambda _{1}c_{12}}+\frac{q_{1}q_{3}}{\lambda _{2}c_{22}}-\frac{%
\left( q_{1}q_{3}+q_{1}q_{4}\right) \widetilde{c}_{2}}{\lambda _{1}+\lambda
_{2}}\right] .  \label{eq26}
\end{equation}
Here, the quantities $q_{i}$ are defined as $q_{1}=p_{11}p_{12}$, $%
q_{2}=p_{21}p_{22}$, $q_{3}=p_{11}p_{22}$, and $q_{4}=p_{12}p_{21}$ with $%
p_{ij}$ specified by the transformation matrix $P$ of Eq. (\ref{eq04}); and
the quantities $\widetilde{c}_{i}$ are $\widetilde{c}_{i}=\left(
1/c_{1i}\right) +\left( 1/c_{2i}\right) $ with $c_{ij}$ defined as $%
c_{ij}=\lambda _{i}\tau ^{\left( j\right) }+1$; for $i$, $j=1$ and $2$.

The stochastic differential equations specified by Eqs. (\ref{eq05-1}) and (%
\ref{eq05-2}) reduce to
\begin{equation}
dy\left( t\right) =F\cdot y\left( t\right) dt+\Lambda \cdot dW\left(
t\right) .  \label{eq27}
\end{equation}%
for the case of white noise. Similarly to the case of colored noise, the
solution of Eq. (\ref{eq27}) can be put in the form of Ito integral, and we
obtain the variances and covariance of $y_{1}$ and $y_{2}$ as \ \
\begin{equation}
\sigma _{1}^{\left( 0\right) 2}=d^{\left( 1\right) }\left[ \frac{q_{3}^{2}}{%
\lambda _{1}}+\frac{q_{4}^{2}}{\lambda _{2}}-\frac{4q_{3}q_{4}}{\lambda
_{1}+\lambda _{2}}\right] +d^{\left( 2\right) }\left[ \frac{q_{1}^{2}}{%
\lambda _{1}}+\frac{q_{1}^{2}}{\lambda _{2}}-\frac{4q_{1}^{2}}{\lambda
_{1}+\lambda _{2}}\right] ,  \label{eq24-1}
\end{equation}%
\begin{equation}
\sigma _{2}^{\left( 0\right) 2}=d^{\left( 1\right) }\left[ \frac{q_{2}^{2}}{%
\lambda _{1}}+\frac{q_{2}^{2}}{\lambda _{2}}-\frac{4q_{2}^{2}}{\lambda
_{1}+\lambda _{2}}\right] +d^{\left( 2\right) }\left[ \frac{q_{3}^{2}}{%
\lambda _{2}}+\frac{q_{4}^{2}}{\lambda _{1}}-\frac{4q_{3}q_{4}}{\lambda
_{1}+\lambda _{2}}\right] ,  \label{eq25-1}
\end{equation}%
and
\begin{equation}
E^{\left( 0\right) }=d^{\left( 1\right) }\left[ \frac{q_{2}q_{3}}{\lambda
_{1}}+\frac{q_{2}q_{4}}{\lambda _{2}}-\frac{2\left(
q_{2}q_{3}+q_{2}q_{4}\right) }{\lambda _{1}+\lambda _{2}}\right] +d^{\left(
2\right) }\left[ \frac{q_{1}q_{4}}{\lambda _{1}c_{12}}+\frac{q_{1}q_{3}}{%
\lambda _{2}c_{22}}-\frac{2\left( q_{1}q_{3}+q_{1}q_{4}\right) }{\lambda
_{1}+\lambda _{2}}\right] ,  \label{eq26-1}
\end{equation}%
where the superscript $\left( 0\right) $ in $\sigma $ and $E$ is used to
denote the result of white noise. These are exactly the same as those given
by Eqs. (\ref{eq24})-(\ref{eq26}) but with $\tau ^{\left( 1\right) }=\tau
^{\left( 2\right) }=0$. Following this, by setting either $\tau ^{\left(
1\right) }=0$ or $\tau ^{\left( 2\right) }=0$ in Eqs. (\ref{eq24})-(\ref%
{eq26}) we have the results for the case of Gaussian colored noise in one
component and white noise in the other.

\section{Results and discussions}

The above results are applied to analyze the fluctuations of two systems,
including auto-regulatory networks and toggle switches. The fluctuations of
the system near a stable point are analyzed by measuring the variance of $%
x_{i}$ in terms of the Fano factors $\nu _{i}$, defined as
\begin{equation}
\nu _{i}\left( \tau ^{\left( 1\right) },\tau ^{\left( 2\right) }\right)
=\Omega \left[ \frac{\sigma _{i}^{2}\left( \tau ^{\left( 1\right) },\tau
^{\left( 2\right) }\right) }{x_{i}^{\ast }}\right] ,  \label{eq30}
\end{equation}%
and the covariance of $x_{1}$ and $x_{2}$ in terms of the correlation
coefficient $R_{12}$, defined as
\begin{equation}
R_{12}\left( \tau ^{\left( 1\right) },\tau ^{\left( 2\right) }\right) =\frac{%
E\left( \tau ^{\left( 1\right) },\tau ^{\left( 2\right) }\right) }{\sigma
_{1}\left( \tau ^{\left( 1\right) },\tau ^{\left( 2\right) }\right) \sigma
_{2}\left( \tau ^{\left( 1\right) },\tau ^{\left( 2\right) }\right) }.
\label{eq31}
\end{equation}%
Here $\sigma _{i}\left( \tau ^{\left( 1\right) },\tau ^{\left( 2\right)
}\right) =\sqrt{\sigma _{i}^{2}\left( \tau ^{\left( 1\right) },\tau ^{\left(
2\right) }\right) }$ is the standard deviation of the concentration $x_{i}$
near a stable equilibrium. Note that we add the superscript, $\left(
0\right) $, to a quantity to refer to the results of white noises, i.e. $%
\tau ^{\left( 1\right) }=0$ and $\tau ^{\left( 2\right) }=0$. The Fano
factor is equal to one, $\nu _{i}=1$, for a Poisson process. Based on this,
we refer a process with Fano factor smaller than one as sub-Poissonian and a
process with Fano factor larger than one as super-Poissonian. Thus, the
sub-Poissonian deviates from the inherent randomness of Poisson process in a
opposite way to the super-Poissonian, the former suppresses the occurrence
probability of the large deviations from the mean value $x^{\ast }$
meanwhile the latter increases it.

\subsection{Auto-regulatory network}

\begin{figure}
\includegraphics[width=0.95\linewidth,angle=0]{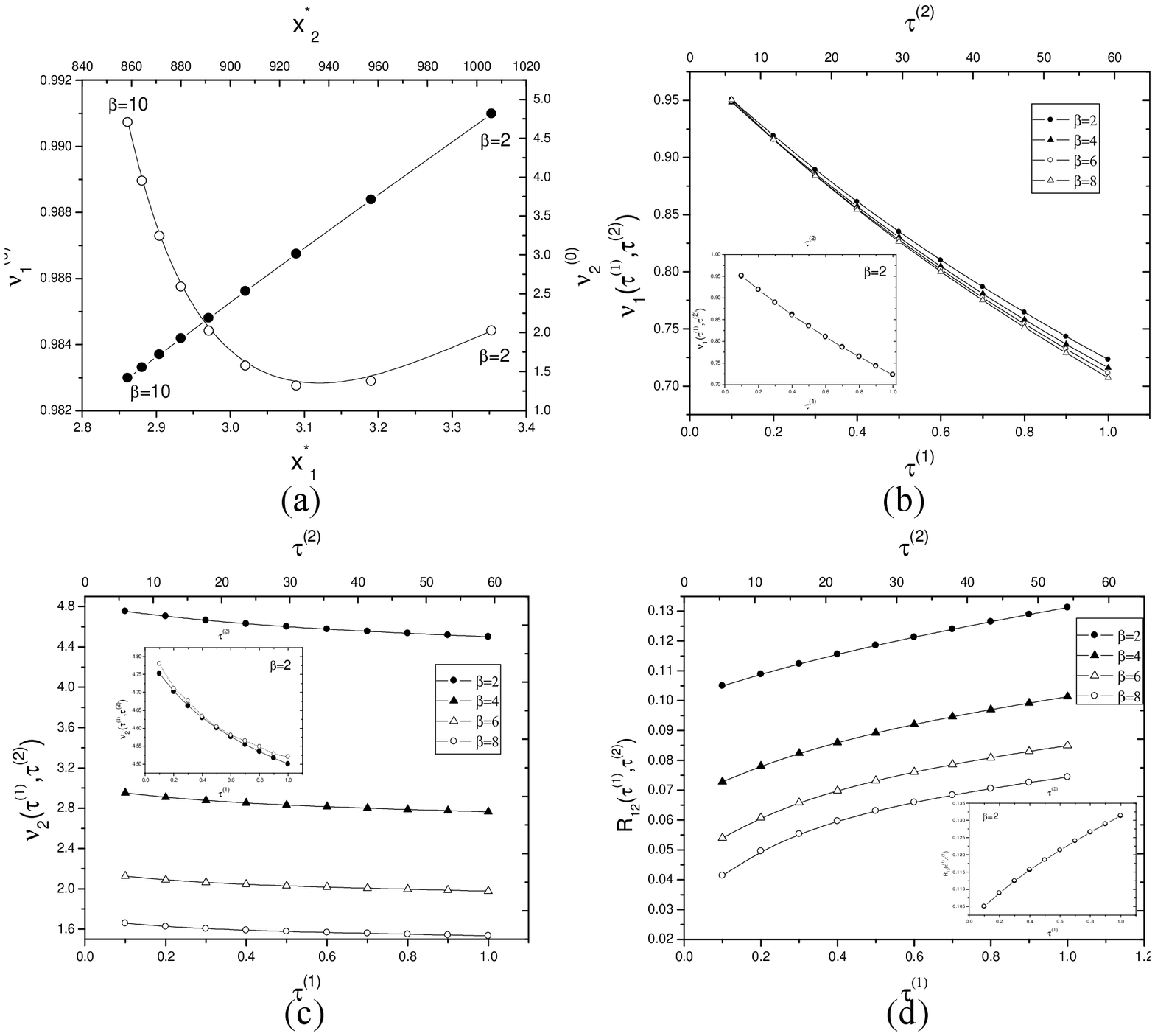}
\hfill \caption{The fluctuations for autoregulatory networks: $\
$For the case of white noise, the Fano factors, $\nu _{1}^{\left(
0\right) }$ (hollow circles, left vertical) and $\nu _{2}^{\left(
0\right) }$ (black dots, right vertical), versus the concentrations
of mRNA (lower horizontal) and protein
(upper horizontal) at the stable points for different Hill coefficients $%
\beta $ are shown in $\left( a\right) $. For the case of Gaussian
colored noise,\ the Fano factors, $\nu _{1}\left( \tau ^{\left(
1\right) },\tau ^{\left( 2\right) }\right) $ and $\nu _{2}\left(
\tau ^{\left( 1\right) },\tau ^{\left( 2\right) }\right) $, and
the correlation coefficients of noises $R_{12}\left( \tau ^{\left(
1\right) },\tau ^{\left( 2\right) }\right) $ versus the
correlation times of noises, $\tau ^{\left( 1\right) }$ (lower
horizontal) and $\tau ^{\left( 2\right) }$ (upper horizontal), at
the stable points for different Hill coefficients $\beta $ are
shown in $\left( b\right) $, $\left( c\right) $, and $\left(
d\right) $, respectively.} \label{fig1}
\end{figure}

\begin{figure}
\includegraphics[width=0.95\linewidth,angle=0]{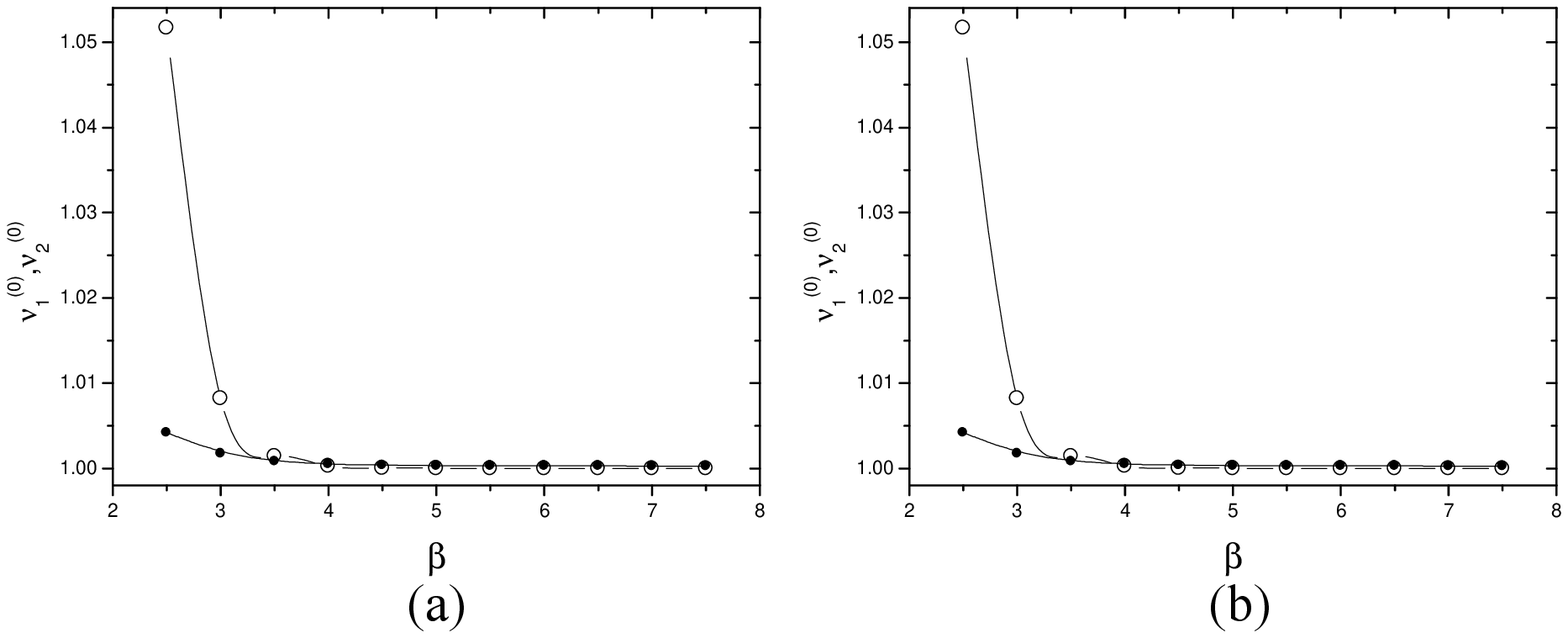}
\hfill \caption{The Fano factors of toggle switch, $\nu
_{1}^{\left( 0\right) }$ (hollow circles) and $\nu _{2}^{\left(
0\right) }$ (black dots), versus the Hill coefficient $\beta $ are
shown in $\left( a\right) $ for the stable point $A$ and in
$\left( b\right) $ for the stable point $B$. Note that while $\nu
_{1}^{\left( 0\right) }$ and $\nu _{2}^{\left( 0\right) }$ are
shown with the same vertical scale in $\left( a\right) $, the left
vertical scale is for $\nu _{1}^{\left( 0\right) }$ and the right
vertical scale is for $\nu _{2}^{\left( 0\right) }$ in $\left(
b\right) $.} \label{fig2}
\end{figure}

The two variables of auto-regulatory network, $x_{1}$ and $x_{2}$, are
referred to the concentrations of mRNA and protein, respectively. For the
functions $R_{i}\left( x\right) $ in the drift force of Eq. (\ref{eq006}),
we adopt the most common noise-attenuating regulatory mechanism, called
negative feedback and described by Hill function,
\begin{equation}
R_{1}\left( x_{2}\right) =\frac{k_{\max }}{1+\left( x_{2}/k_{d}\right)
^{\beta }},  \label{eq28}
\end{equation}%
to regulate the production of mRNA, and set
\begin{equation}
R_{2}\left( x_{1}\right) =k_{2}x_{1}.  \label{eq29}
\end{equation}%
Here $k_{\max }$ is the maximum transcription rate of mRNA, $k_{d}$ is the
binding constant specifying the threshold protein concentration at which the
transcription rate is half its maximum value, $\beta $ is the Hill
coefficient, and $k_{2}$ is the translation rate of protein $k_{2}$. Then, a
stable equilibrium, $x^{\ast }$, can be characterized by two conditions: $\
\theta ^{\left( 1\right) }\theta ^{\left( 2\right) }-r_{1}\left( x_{2}^{\ast
}\right) k_{2}>0$ and $\theta ^{\left( 1\right) }+\theta ^{\left( 2\right)
}>0$ with $r_{1}\left( x_{2}^{\ast }\right) =\left. \partial R_{1}\left(
x_{2}\right) /\partial x_{2}\right\vert _{x_{2}=x_{2}^{\ast }}$.
Subsequently, one can use Bendixson's criterion to further conclude that
there are no cycles and only one equilibrium exists \cite{bendixson}.

We mainly follow Refs. \cite{thattai,tao-c} to specify the values of the
parameters as follows. The half-lifetimes of mRNA molecules and proteins are
set as $2$ minutes and $1$ hour, respectively; this leads to $\theta
^{\left( 1\right) }=\left( \ln 2\right) /2$ and $\theta ^{\left( 2\right)
}=\left( \ln 2\right) /60$ in the unit of $\left( \min \right) ^{-1}$.\ The
average size of a burst of proteins, $b=k_{2}/\theta ^{\left( 1\right) }$,
is set as $10$. This leads to $k_{2}=5\left( \ln 2\right) $. By using the
fact that the protein concentration is about $1200$ when $\beta =0$ (no
feedback), we set $k_{\max }=3$ \cite{remark}. To study the effect of the
strength of negative feedback on the fluctuation of the system, we vary the
parameters $\beta $ from $2$ to $10$, while the $k_{d}$ value is fixed as $%
800$.

We first consider the case of white noise. Based on Eqs. (\ref{eq24-1})-(%
\ref{eq26-1}), we obtain the $\nu _{1}^{\left( 0\right) }$and $\nu
_{2}^{\left( 0\right) }$ values for $\beta $ ranging from $2$ to $10$, and
the results are shown in Fig. 1(a). The characteristic features revealed
from the results are summarized as follows. The $\nu _{2}^{\left( 0\right) }$
value is much larger than the corresponding $\nu _{1}^{\left( 0\right) }$
value, and it decreases as $\beta $ increases. This leads to the well-known
conclusions that the stochastic fluctuations occur mainly at the translation
level, and the negative feedback may enhance the stability of the system. On
the other hand, the $\nu _{1}^{\left( 0\right) }$ value decreases from $\nu
_{1}^{\left( 0\right) }=0.9844$ at $\beta =2$, reaches the minimum $\nu
_{1}^{\left( 0\right) }=0.9827$ at $\beta =4$, and then increases to $\nu
_{1}^{\left( 0\right) }=0.9907$ at $\beta =10$. All the $\nu _{1}^{\left(
0\right) }$ values are very close to but always less than one. This implies
that the process of transcription is sub-Poissonian. We then consider the
effect of non-zero correlation time of noise on the fluctuation of the
system based on\ Eqs. (\ref{eq24})-(\ref{eq26}). First, we show the plots
of $\nu _{1}\left( \tau ^{\left( 1\right) },\tau ^{\left( 2\right) }\right) $
and $\nu _{2}\left( \tau ^{\left( 1\right) },\tau ^{\left( 2\right) }\right)
$ versus $\beta $ for $\left( \tau ^{\left( 1\right) },\tau ^{\left(
2\right) }\right) =\left( 0.1,6\right) $, $\left( 0.2,12\right) $, $\left(
0.4,24\right) $, and $\left( 0.6,36\right) $ in Figs. 1(b) and 1(c). Here
the values of $\left( \tau ^{\left( 1\right) },\tau ^{\left( 2\right)
}\right) $ are set as the multiples $10$, $20$, $40$, and $60$ of the
half-lifetimes of mRNA and protein, respectively. The results indicate that
the fluctuations are reduced by the amount proportional to the
correlation time value of the noise. However, the reduction is less significant for
protein owing to its long half-lifetime when compared with that of mRNA.
Moreover, the effect of reduction, in general, is enhanced for a larger Hill
coefficient, but this is not very significant for mRNA owing to the nature
of the sub-Poissonian process. We also show the effect of the correlation time
of noise on the correlation coefficient between $x_{1}$ and $x_{2}$ in Fig.
1(d), where the plots of $R_{12}\left( \tau ^{\left( 1\right) },\tau
^{\left( 2\right) }\right) $ versus $\beta $ for five sets of $\left( \tau
^{\left( 1\right) },\tau ^{\left( 2\right) }\right) $ values are displayed.
The results indicate that the correlation coefficient $R_{12}$ increases
with the correlation times of the noises but decreases with the Hill coefficient.
Moreover, as the consequence of sub-Poissonian distribution for the
mRNA component, the $R_{12}$ values between two components are very
small, ranged between $0.0202$ and $0.1213$.

The analytical results given by Eqs. (\ref{eq24})-(\ref{eq26}) are obtained
with the approximation of linearization about a stable equlibrium. To test
the validity of such an approximation, we solve the stochastic differential
equations numerically by using Heun's method. This numerical method is a
stochastic version of the Euler method, which reduces to the second-order
Runge-Kutta method in the absence of noise \cite{toral}. The numerical
simulations are in a very good agreement with the analytical results as shown
in Figs. 1(b)-1(d) for $\beta = 2.0$.

\subsection{Toggle switch}

\begin{figure}
\includegraphics[width=0.95\linewidth,angle=0]{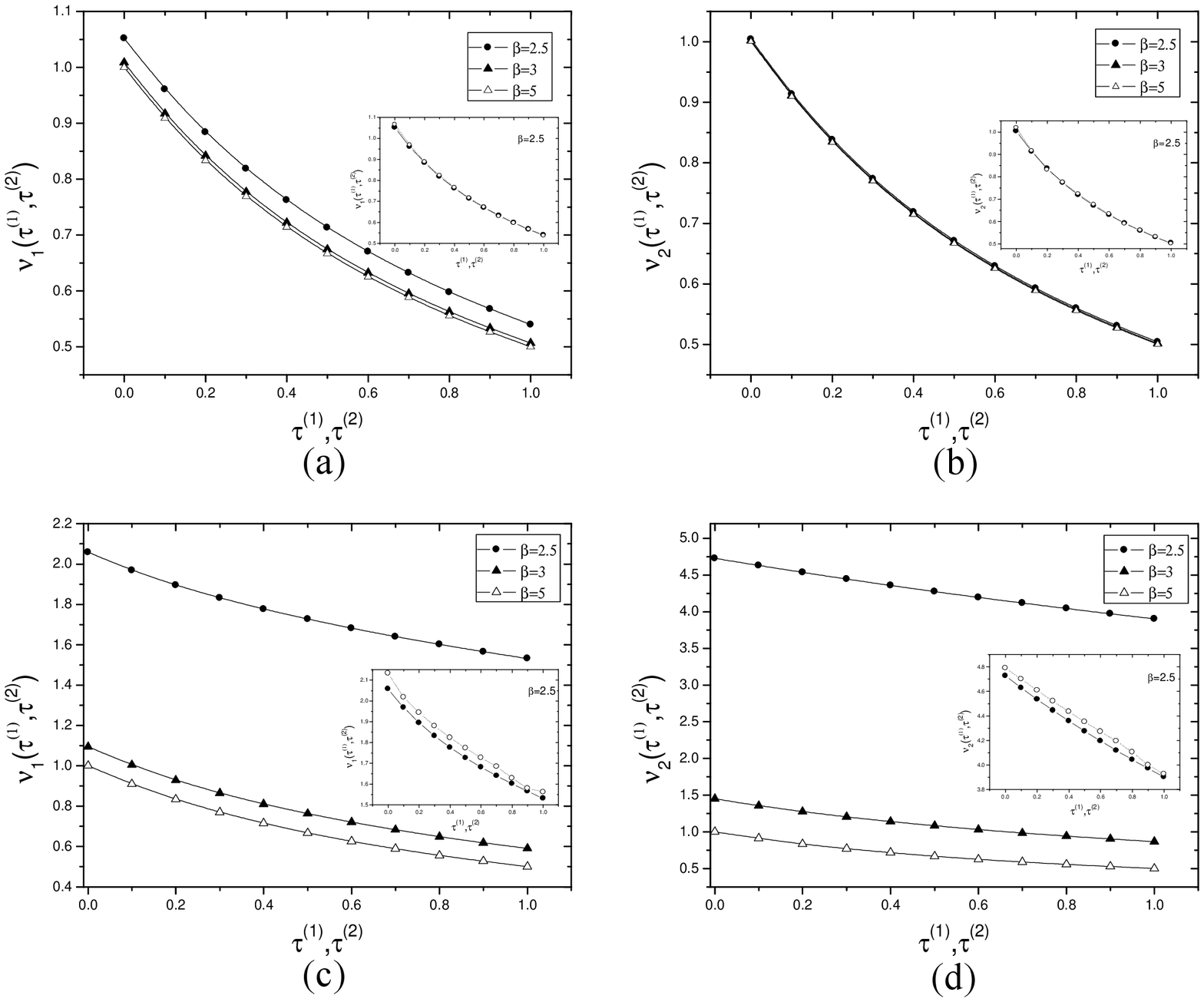}
\hfill \caption{The fluctuations of toggle switch for the case of
Gaussian colored noise: The Fano factors, $\nu _{1}\left( \tau
^{\left( 1\right) },\tau ^{\left( 2\right) }\right) $ and $\nu
_{2}\left( \tau ^{\left( 1\right) },\tau ^{\left( 2\right)
}\right) $, versus the correlation times of noises, $\tau ^{\left(
1\right) }$ and $\tau ^{\left( 2\right) }$ for different Hill
coefficients $\beta $ are shown in $\left( a\right) $ and $\left(
b\right) $ for the stable point $A$ and $\left( c\right) $ and
$\left( d\right) $ for the stable point $B$. Note that $\tau
^{\left( 1\right) }$ is set to be
equal to $\tau ^{\left( 2\right) }$ in the calculations of $\nu _{1}$ and $%
\nu _{2}$, and the values of $\nu _{1}$ and $\nu _{2}$ for $\tau
^{\left( 1\right) }=\tau ^{\left( 2\right) }=0$ are the results
for the case of white noise.} \label{fig3}
\end{figure}

\begin{figure}
\includegraphics[width=0.95\linewidth,angle=0]{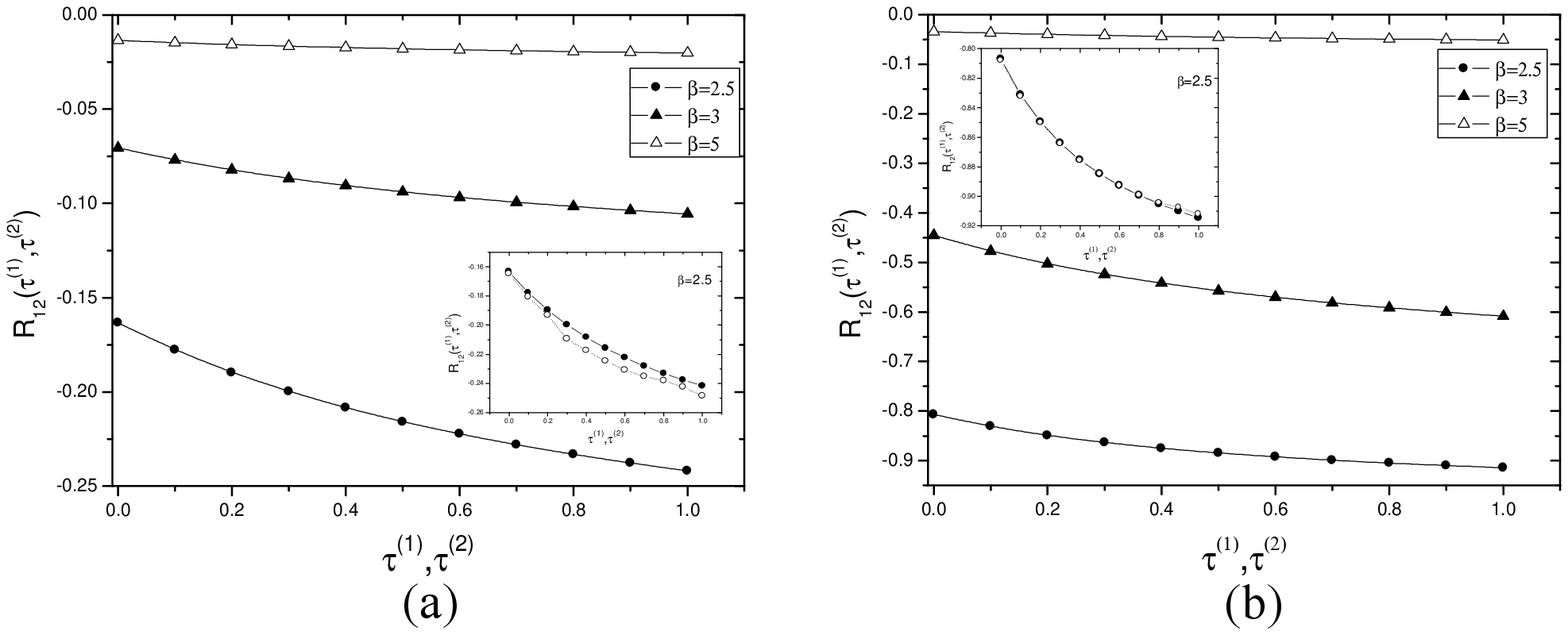}
\hfill \caption{The correlation coefficients of noises of toggle switch, $%
R_{12}\left( \tau ^{\left( 1\right) },\tau ^{\left( 2\right)
}\right) $, versus the correlation times of noises, $\tau ^{\left(
1\right) }$ and $\tau
^{\left( 2\right) }$ for different Hill coefficients $\beta $ are shown in $%
\left( a\right) $ for the stable point $A$ and in $\left( b\right)
$ for the
stable point $B$. Here $\tau ^{\left( 1\right) }$ is set to be equal to $%
\tau ^{\left( 2\right) }$, and the results for $\tau ^{\left(
1\right) }=\tau ^{\left( 2\right) }=0$ correspond to the case of
white noise.} \label{fig4}
\end{figure}

A toggle switch consists of two transcription factors with concentrations $%
x_{1}$ and $x_{2}$; the transcription factors can regulate each other's
synthesis through the negative feedback mechanism. Such genetic circuits
often exhibit two stable states, referred as $A\left( x_{1}^{\ast
},x_{2}^{\ast }\right) $ and $B\left( x_{1}^{\ast },x_{2}^{\ast }\right) $,
for which, the $x_{1}$ component is dominant with almost vanishing $x_{2}$
component for the state $A$, and vice versa for the state $B$. The distinct
two stable states can be switched either spontaneously or by a driven
signal. A plasmid of this type in \textit{Escherichia coli} has been
engineered by Gardner, Cantor, and Collins \cite{gardner}. For this,
correspondingly, the degradation rates are rescaled to $\theta ^{\left(
1\right) }=\theta ^{\left( 2\right) }=1$; the regulation functions in the
drift force of Eq. (\ref{eq006}) then become
\begin{equation}
R_{1}\left( x_{2}\right) =\frac{\alpha _{1}}{1+x_{2}^{\beta }}  \label{eq33}
\end{equation}%
and
\begin{equation}
R_{2}\left( x_{1}\right) =\frac{\alpha _{2}}{1+x_{1}^{\gamma }}.
\label{eq34}
\end{equation}%
Here the parameter values are $\alpha _{1}=156.25$, $\alpha _{2}=15.6$, $%
\beta =2.5$, and $\gamma =1.0$. Such a specification leads to two stable
states, $A\left( 155.7634,0.0995\right) $ and $B\left( 0.3324,11.7080\right)$.

The system with the regulation functions of Eqs. (\ref{eq33}) and (\ref{eq34})
exhibits the bistability over a wide rage of parameter values. In this
study, we intend to analyze how the characteristics of the fluctuation
change with the cooperative binding of the system. Thus, we calculate the
variances and covariances along the parametric path increasing the Hill
coefficient $\beta $ from $2.5$ to $7.5$ and keeping the other parameters
to be the same as the previous values. As the $\beta$ value varied from $2.5$ to $7.5$,
the loci of the two stable states are changed as follows. The $%
x_{1}^{\ast} $ value increases slightly from $155.7634$ up to $156.2500$ and
the $x_{2}^{\ast }$ value decreases insignificantly from $0.0995$ down to $%
0.0992$ for the stable state $A$; meanwhile, the $x_{1}^{\ast }$ value
decreases from $0.3324$ down to $1.759 \cdot 10^{-7}$ and the $x_{2}^{\ast }$
value increases from $11.7080$ up to $15.6000$ for the stable state $B$.
Accordingly, increasing the $\beta $ value will enhance the major component
and suppress the minor component, and it will shift the location of the
stable state $B$ more significantly than that of $A$.

We first consider the case of white noise for the study of fluctuations. The
results of $\nu _{1}^{\left( 0\right) }$and $\nu _{2}^{\left( 0\right) }$
for different $\beta $ values are shown in Figs. 2(a) and 2(b) for the
stable states $A$ and $B$, respectively. As indicated by the numerical
results, the $\nu _{1}^{\left( 0\right) }$and $\nu _{2}^{\left( 0\right) }$
values all are larger than one; thus, the fluctuations are caused by
super-Poissonian processes for both $A$ and $B$. This is opposite to the
case of the auto-regulatory network, and it agrees with the results obtained by
Tao \cite{tao}. Moreover, the system in the stable state $B$ always
possesses a larger deviation from Poissonian than that for $A$. But, the
maximum deviation occurs at $\beta =2.5$ for both $A$ and $B$ at which, we
have $\nu _{1}^{\left( 0\right) }=1.0517$ and $\nu _{2}^{\left( 0\right)
}=1.0042$ for state $A$ and $\nu _{1}^{\left( 0\right) }=2.0580$ and $\nu
_{2}^{\left( 0\right) }=4.7268$ for state $B$. Note that there is a big drop
in the $\nu _{1}^{\left( 0\right) }$and $\nu _{2}^{\left( 0\right) }$ values
for the system in the state $B$ when the $\beta $ value increases from $2.5$
to $3.0$. As the values of the finite correlation time of the noises set in,
the resultant values of $\nu _{1}\left( \tau ^{\left( 1\right) },\tau ^{\left( 2\right) }\right)$
and $\nu _{2}\left( \tau ^{\left( 1\right) },\tau ^{\left( 2\right) }\right)
$ for the values of $\tau ^{\left( 1\right) }=\tau ^{\left( 2\right) }$
ranging from $0$ to $1$ with $\beta =2.5$, $3$, and $6$ are shown in Figs.
3(a) and 3(b) for the states $A$ and in Figs. 3(c) and 3(d) for the state $B$.
The results indicate a similar feature as that for the case of the
auto-regulatory network, namely the fluctuations are reduced by the noises
being correlated, and the longer is the correlation time, the bigger amount
the fluctuation decreases. We also show the plots of $R_{12}\left( \tau
^{\left( 1\right) },\tau ^{\left( 2\right) }\right) $ versus $\tau ^{\left(
1\right) }\left( =\tau ^{\left( 2\right) }\right) $ for three different $%
\beta $ values in Figs. 4(a) and 4(b) for the stable states $A$ and $B$,
respectively. All $R_{12}\left( \tau ^{\left( 1\right) },\tau ^{\left(
2\right) }\right) $ values shown in the figures are negative. This implies
that the fluctuations of two components are anti-correlated, which reflects
the fact that the two components are negatively regulated with each other.
In particular, the system in the state $B$ with $\beta =2.5$ and $3.0$ is
highly anti-correlated with $R_{12}^{\left( 0\right) }=-0.8070$ and $-0.4453$,
respectively. Also, the anti-correlation of fluctuations is enhanced as
the finite correlation times of noises are set in.

The results shown in the above are also compared with those obtained from
numerical simulations. The comparison is shown in the insets of Figs.
3(a)-3(d) and Figs. 4(a)-4(b) for $\beta =2.5$, and much larger differences
in the Fano factors have been observed for the stable point $B$ than that
for the stable point $A.$

\section{Summary}

Noise due to stochastic fluctuations is always present and essential in the
gene expression process due to low copy numbers of molecules and stochastic
nature of biochemical reactions. A framework has been set up for the study of the
noise by constructing the equivalent Langevin description of the system.
By taking then the fluctuation as Gaussian colored noise, we first
solve the Langevin equation, and then obtain the general formulae for the
variances and covariance of a two-dimensional system near a stable
equilibrium based on Ito calculus. For the latter, two important lemmas,
concerning the variance of a Ito integral and the covariance between two Ito
integrals, are established.

We apply the general formulae to the auto-regulation network and the toggle switch
for the study of stochastic fluctuations of the systems, in particular, the
effect caused by finite correlation time. In general, as correlation time of
noise is set in, it will decrease the fluctuation and enhance the
correlation of fluctuations. For the auto-regulation network, the
fluctuations of mRNA concentration mirror a sub-Poissonian process
with the Fano factor very close to but less than one. Consequently, this
leads to the correlation coefficient between the fluctuations of mRNA and
protein concentrations to be very small. For the toggle switch it is
peculiar for one of the two stable states when the system has the Hill
coefficient $\beta $ being around $2.5$; the fluctuations are caused by a process
far away from Poissonian, and the fluctuations between two molecular
components are highly anti-correlated. Aside from this, the fluctuations are
caused by super-Poissonian processes with Fano factors very close to but
larger than one for both concentrations in any of the two stable states; and
this then leads to small but negative correlation coefficients between the
fluctuations of two components. Moreover, as it has been shown, these
analytical results are in a very good agreement with the numerical simulations.

To summarize, let us stress that it is more realistic to take into account
the noise finite correlation time in order to study of stochastic fluctuations
of gene regulatory networks. Meanwhile, in this work the noises of two components
are still assumed to be independent, although the components couple with each other
in the rate equations. It was shown \cite{cao} that increasing the coupling
strength between two noises may drive the system to transit from a bistable
stationary probability distribution to a mono-stable one. Thus, it would be
interesting to study the effect of coupled noises on the stochastic fluctuations
in the gene regulatory networks. Besides, it will be also very interesting to extend
the work to the case of a non-Gaussian colored noise \cite{bph}.

We thank B.C. Bag, C.-K. Hu, and D. Salahub for discussions. The work of
M.C.H., J.W.W., and Y.P.L. was partially supported by the National Science
Council of Republic of China (Taiwan) under the Grant No. NSC
96-2112-M-033-006. K.G.P. was supported by Grants NSC 96-2811-M-001-018 and
NSC 97-2811-M-001-055.


\begin{thebibliography}{99}
\bibitem{alon} U. Alon, \textit{An Introduction to Systems Biology: Design
Principles of Biological Circuits} (CRC Press, 2006).

\bibitem{ptashne} M. Ptashne, \textit{A Genetic Switch: Phage $\lambda$ and
Higher Organisms}, 2nd ed. (Cell Press and Blackwell Scientific, Cambridge,
MA, 1992).

\bibitem{hasty} J. Hasty, J. Pradines, M. Dolnik, and J.J. Collins, Proc.
Natl. Acad. Sci. \textbf{97}, 2075 (2000).

\bibitem{gardner} T.S. Gardner, C.R. Cantor, and J.J. Collins, Nature
\textbf{403}, 339 (2000).

\bibitem{elowitz} M.B. Elowitz and S. Leibler, Nature \textbf{403}, 335
(2000).

\bibitem{atkinson} M.R. Atkinson, M.A. Savageau, J.T. Myers, and A.J. Ninfa,
Cell \textbf{113}, 597 (2003).

\bibitem{rao} C.V. Rao, D.M. Wolf, and A.P. Arkin, Nature \textbf{420}, 231
(2002).

\bibitem{pedraza}J.M. Pedraza and J. Paulsson, Science {\bf 319}, 339 (2008).

\bibitem{becskei} A. Becskei and L. Serrano, Nature \textbf{405}, 590 (2000).

\bibitem{thattai} M. Thattai and A. van Oudenaarden, Proc. Natl. Acad. Sci.
\textbf{98}, 8614 (2001).

\bibitem{ozbudak} E.M. Ozbudak, M. Thattai, I. Kurtser, A.D. Grossman, and
A. van Oudenaarden, Nature Genetics \textbf{31}, 69 (2002).

\bibitem{cherry} J.L. Cherry and F.R. Adler, J. Theor. Biol. \textbf{203},
117 (2000).

\bibitem{warren} P.B. Warren and P.R. ten Wolde, Phys. Rev. Lett. \textbf{92}, 128101 (2004);
J. Phys. Chem. B \textbf{109}, 6812 (2005).

\bibitem{loinger} A. Lipshtat, A. Loinger, N.Q. Balaban, and O. Biham, Phys.
Rev. Lett. \textbf{96}, 188101 (2006); A. Loinger, A. Lipshtat, N.Q.
Balaban, and O. Biham, Phys. Rev. E \textbf{75}, 021904 (2007).

\bibitem{gillespie} D.T. Gillespie, J. Chem. Phys. \textbf{81}, 2340 (1977).

\bibitem{arkin} A. Arkin, J. Ross, and H.H. McAdams, Genetics \textbf{149},
1633 (1998).

\bibitem{kampen} N.G. van Kampen, \textit{Stochastic Processes in Physics
and Chemistry} (North-Holland, Amsterdam, 1992).

\bibitem{tao} Y. Tao, J. Theor. Biol. \textbf{229}, 147 (2004); \textbf{231}, 563 (2004).

\bibitem{tao-c} Y. Tao, Y. Jia, and T.G. Dewey, J. Chem. Phys. \textbf{122},
124108 (2005).

\bibitem{jung} P. Jung and P. H\"{a}nggi, Phys. Rev. A \textbf{35}, 4464
(1987); J. Opt. Soc. Am. B \textbf{5}, 979 (1988).

\bibitem{shahrezaei} V. Shahrezaei, J.F. Ollivier, and P.S. Swain, Molecular
Systems Biology \textbf{4:196} (2008).

\bibitem{gardiner} C.W. Gardiner, \textit{Handbook of Stochastic Methods},
3rd ed. (Springer-Verlag, Berlin, 2004).

\bibitem{uhlenbeck} G.E. Uhlenbeck and L.S. Ornstein, Phys. Rev. \textbf{36}%
, 823 (1930); M.C. Wang and G.E. Uhlenbeck, Rev. Mod. Phys. \textbf{17}, 323
(1945).

\bibitem{bendixson} E. Coddington and N. Levinson, \textit{Theory of
Ordinary Differential Equations} (McGrow-Hill, N.Y., 1955).

\bibitem{remark} The $k_{\max }$ value was set as $2\ln 2\simeq 1.4$ in Ref.
\cite{thattai}, this corresponds to a mean protein number $600$ when $\beta
=0$ and leads to the increase of mean protein number as the Hill coefficient
$\beta $ increases. Our setting, $k_{\max }=3$ which is the same as that
given in Ref. \cite{tao-c}, leads to the decrease of mean protein number as $%
\beta $ increases. However, different settings do not change the picture
obtained in this work.

\bibitem{toral} R. Toral, in \textit{Computational Physics}, Lecture Notes
in Physics Vol. 448, edited by P. Garrido and J. Marro (Springer-Verlag,
Berlin, 1995).

\bibitem{cao} L. Cao, D.J. Wu, and S.Z. Ke, Phys. Rev. E \textbf{52}, 3228
(1995).

\bibitem{bph} B.C. Bag, K.G. Petrosyan, and C.-K. Hu, Phys. Rev. E \textbf{76}, 056210 (2007).

\end{thebibliography}
\end{document}